\documentclass[%
 preprint,
 superscriptaddress, numerical
 amsmath,amssymb,
 aps, prl
]{revtex4-1}

\usepackage{graphicx}
\usepackage{dcolumn}
\usepackage{bm}
\usepackage{mathtools}

\begin{document}


\title{Current-induced spin polarization in InGaAs and GaAs epilayers with varying doping densities}

\author{M. Luengo-Kovac}
\affiliation{ 
Department of Physics, University of Michigan
}

\author{S. Huang}
\affiliation{
Department of Materials Science and Engineering, University of Michigan
}

\author{D. Del Gaudio}
\affiliation{
Department of Materials Science and Engineering, University of Michigan
}

\author{J. Occena}
\affiliation{
Department of Materials Science and Engineering, University of Michigan
}

\author{R. S. Goldman}
\affiliation{
Department of Materials Science and Engineering, University of Michigan
}

\author{R. Raimondi}
\affiliation{
Dipartimento di Matematica e Fisica, Universit\`{a} Roma Tre
}

\author{V. Sih}
\affiliation{
Department of Physics, University of Michigan
}


\begin{abstract}
The current-induced spin polarization and momentum-dependent spin-orbit field were measured in In$_{x}$Ga$_{1-x}$As epilayers with varying indium concentrations and silicon doping densities. Samples with higher indium concentrations and carrier concentrations and lower mobilities were found to have larger electrical spin generation efficiencies. Furthermore, current-induced spin polarization was detected in GaAs epilayers despite the absence of measurable spin-orbit fields, indicating that the extrinsic contributions to the spin polarization mechanism must be considered. Theoretical calculations based on a model that includes extrinsic contributions to the spin dephasing and the spin Hall effect, in addition to the intrinsic Rashba and Dresselhaus spin-orbit coupling, are found to qualitatively agree with the experimental results. 
\end{abstract}
\maketitle


Current-induced spin polarization (CISP), also known as the inverse spin galvanic effect, is a phenomenon in which a bulk electron spin polarization is generated by an electric field applied in the plane of the sample. It has been measured in semiconductor epilayers \cite{Kato_2004, Norman_2014} and in 2-dimensional electron gases (2DEGs) \cite{Sih_2005, Yang_2006}, and is of interest for the development of an all-electrical, all-semiconductor spintronic device \cite{Awschalom_2009}. Indeed, all-electrical spin generation and spin manipulation has been demonstrated in n-InGaAs \cite{Stepanov_2014}. 

However, the polarization mechanism is still unclear. Although it was predicted that the spin polarization should be proportional to the spin-orbit (SO) splitting \cite{Aronov_1991}, measurements performed on InGaAs epilayers showed that the crystal axis with the smallest SO splitting had the largest CISP and vice versa \cite{Norman_2014}. In addition, CISP has been measured in GaN \cite{Koehl_2009} and ZnSe \cite{Stern_2006}, which have weak SO coupling.   

At the origin of the spin polarization by an electric current, there is a lowering of the allowed symmetry transformations. The reduced symmetry implies the appearance of terms linear in momentum in the effective Hamiltonian for the electricity carriers. These linear-in-momentum terms may have both intrinsic or extrinsic character. In the former case, they appear in the effective band Hamiltonian. Such a situation has been studied first  for the Dresselhaus spin-orbit coupling \cite{Aronov_1989} and the Rashba spin-orbit coupling \cite{Edelstein_1990}. Later, both the above spin-orbit couplings were considered \cite{Trushin_2007}, as well as the interplay with a Zeeman term \cite{Engel_2007, Gorini_2010}. In the latter case of extrinsic character, the linear-in-momentum terms appear in the scattering potential \cite{Tarasenko_2007}. 

In this article, we report on measurements of CISP and SO fields in InGaAs and GaAs epilayers with varying indium concentrations and doping densities. The observation of CISP in GaAs epilayers, in which the SO fields are smaller than what we can measure ($<$0.1 mT), suggests that extrinsic mechanisms must be considered in order to explain CISP. We compare our experimental results for InGaAs epilayers to a model proposed by Gorini \textit{et al.} \cite{Gorini_2017} for a 2DEG, which includes intrinsic and extrinsic contributions to the spin dephasing and the spin Hall effect, as well as the inverse spin galvanic effect.

Five InGaAs and two GaAs samples were studied, each consisting of a 500 nm epilayer grown by molecular beam epitaxy (MBE) on a semi-insulating (001) GaAs substrate. All samples were Si-doped at different concentrations. The samples were etched into a cross-shaped channel with arms along the [110] and [1$\overline{1}$0] crystal axes. This allows for the application of an electric field along an arbitrary in-plane crystal axis \cite{Norman_2014}.

Table 1 shows a summary of sample parameters. The indium concentrations are determined from X-ray rocking curves (XRC), which also show the epilayers to be pseudomorphic or nearly pseudomorphic with the substrate, i.e. the strain relaxation is minimal. The carrier concentrations are determined from Hall and van der Pauw measurements performed on the cross-shaped channels. The mobility and SO coefficients $\alpha$ and $\beta$, defined below, are determined from spin-drag measurements \cite{Norman_2010}, and the spin dephasing time $T_2^*$ is determined from time-resolved Faraday rotation (TRFR) measurements. All values are measured at 30 K. 

Spin-orbit coupling in semiconductors manifests as an effective internal magnetic field. In zinc-blende semiconductors, this is described by the Hamiltonian \cite{Meier_2007}
\begin{equation}\label{SOeq}
H_{\text{SO}} = \alpha(k_y\sigma_x - k_x\sigma_y) + \beta(k_y\sigma_x + k_x\sigma_y)
\end{equation}
for x $\| [1\overline{1}0]$ and y $\| [110]$,  where $\alpha$ includes Rashba-like contributions from structural inversion asymmetry and uniaxial strain, and $\beta$ includes linear Dresselhaus-like contributions from bulk inversion asymmetry and biaxial strain \cite{Norman_2010}. As these two components of the SO field have different crystal axis dependences, the anisotropy of the SO field is characterized by the parameter $r = \alpha/\beta$. In our InGaAs samples, the maximum SO field is along [110] and minimum along [1$\overline{1}$0] crystal axes. 

The SO fields are measured by performing pump-probe spin drag measurements on the samples \cite{Kato_2004_Nature}. The samples are mounted on the cold-finger of a continuous flow cryostat, and all measurements are performed at 30 K unless otherwise noted. A tunable-wavelength pulsed Ti:Sapph laser is split into pump and probe pulses, and the relative time delay of the two pulses can be varied using a mechanical delay line. The pump pulse is circularly polarized in order to induce a spin polarization in the sample according to the optical selection rules. The Faraday (Kerr) angle of the transmitted (reflected) linearly polarized probe is measured with a Wollaston prism and balanced photodiode bridge. The InGaAs (GaAs) samples are measured in a transmission (reflection) geometry. Transmission measurements are not possible in the GaAs samples as the wavelength used to probe the epilayer is absorbed by the substrate. The pump and probe are modulated by a photoelastic modulator and optical chopper respectively in order to allow for cascaded lock-in detection. An electromagnet allows for the application of an external magnetic field in the plane of the sample. 

\begin{figure}
\includegraphics[width=8.5cm]{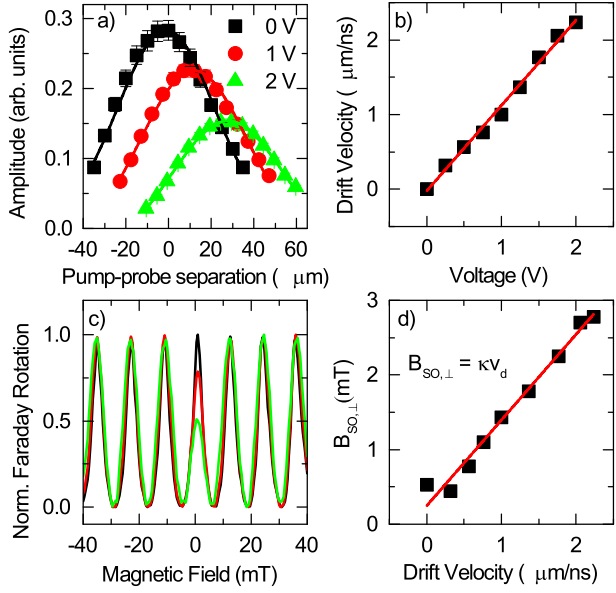}
\caption{\label{fig1} Spin drag measurements for the determination of the SO field for Sample C. (a)  Amplitude $A_0(x)$ vs. pump-probe spatial separation for 0 V (black), 1 V (red) and 2 V (green) and pump-probe time delay $\Delta t = 13$ ns. The location of the center gives the drift velocity. (b) Drift velocity vs. applied voltage. (c) Faraday rotation vs. magnetic field for the same in-plane voltages as (a) at the center of the spin packet. Fits to Eq. \ref{SpinDrageq} give the SO field. (d) The perpendicular component of the SO field at the center of the spin packet vs. drift velocity. The slope $\kappa$ gives the strength of the SO field.}
\end{figure}

When an electric field is applied across the sample, the electron spins precess about the vector sum of the external and SO fields. The Faraday/Kerr rotation $\theta_{F,K}$ can be described by the equation
\begin{equation}\label{SpinDrageq}
\theta_{F,K}(\vec{B}_{\text{ext}},x) = \sum_{n} A_n(x) \times \cos\left[\frac{g\mu_B}{\hbar} \left\vert \vec{B}_{\text{ext}} +\vec{B}_{\text{int}}\right\vert (\Delta t + nt_{\text{rep}})\right]
\end{equation}
where $A_n(x)$ is the amplitude due to successive pump pulses, $g$ is the electron g-factor, $\mu_B$ is the Bohr magneton, $\vec{B}_{\text{ext}}$ is the external magnetic field, $\vec{B}_{\text{int}}$ is the internal SO field, $\Delta t$ is the time delay between the pump and probe pulses, and $t_{\text{rep}} = 13.16$ ns is the time between laser pulses.

\begin{figure}
\includegraphics[width=8.5cm]{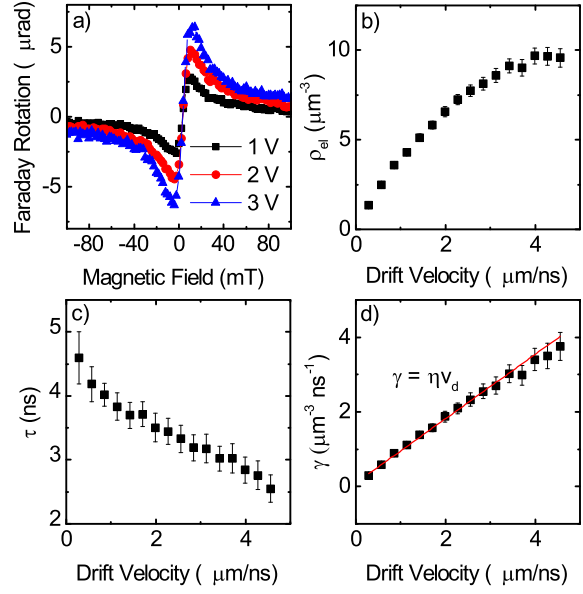}
\caption{\label{fig2} (a) CISP measurements for 1V (black), 2V (red) and 3V (blue), showing an odd-Lorentzian lineshape for Sample A. The spin density $\rho_{\text{el}}$ (b) and lifetime $\tau$ (c) are used to calculate the spin generation rate $\gamma$ (c). The slope  $\eta$ of $\gamma$ with respect to the drift velocity is used to characterize the strength of the CISP.}
\end{figure}

\begin{table*}
\hspace*{-1.5cm}
\begin{tabular}[b]{|c|c|c|c|c|c|c|c|c|}
\hline
Sample & x$_{\text{In}}$ & n  & $\mu$  &  $\frac{m^*}{\hbar}\alpha$ & $\frac{m^*}{\hbar}\beta$  & r & $T_2^*$ & $\rho_{\text{el}}/\theta_{\text{el}}$ ([1$\overline{1}$0],[110])\\
 & & (10$^{16}$ cm$^{-3}$) & (cm$^2$/Vs) &  (neV ns/$\mu$m) &  (neV ns/$\mu$m) &  &  (ns) &  $(\mu \text{m}^{-3}/\mu\text{rad})$ \\ \hline
A & 0.026 & 20.8 $\pm$ 0.1 & 3200 $\pm$ 200 & 26 $\pm$ 5  & 27 $\pm$ 5 & 1.0 $\pm$ 0.3 & 5.1 $\pm$ 0.2 & 0.46, 0.73\\ \hline
B & 0.026 & 15.5 $\pm$ 0.6 & 3400 $\pm$ 300 & 39 $\pm$ 17 & 5.7 $\pm$ 17 & 6.9 $\pm$ 21 & 5.58 $\pm$ 0.07 & 1.42, 1.40\\ \hline
C & 0.024 & 1.58 $\pm$ 0.03 & 6500 $\pm$ 200 & 28 $\pm$ 13 & 2.9 $\pm$ 13 & 9.8 $\pm$ 43 & 7.67 $\pm$ 0.08 & 0.24, 0.27 \\ \hline
 D & 0.02 & 2.93 $\pm$ 0.04  & 5100 $\pm$ 300 & -4.2 $\pm$ 16 & 28 $\pm$ 16 & 0.15 $\pm$ 0.6 & 17.9 $\pm$ 0.2  & 0.023, 0.022 \\ \hline
E & 0.02 & 0.270 $\pm$ 0.002  & 6600 $\pm$ 500 & 13 $\pm$ 4 & 22 $\pm$ 4 & 0.61 $\pm$ 0.2 & 7.3 $\pm$ 0.1 & 0.0043, 0.0043  \\ \hline
F & 0.0 & 51.2 $\pm$ 0.2 & 2600 $\pm$ 200 & -  & - &  - & 6.8 $\pm$ 0.1 & - \\ \hline
G & 0.0 & 3.00 $\pm$ 0.03  & 4600 $\pm$ 100 & - & - &  - & 3.87 $\pm$ 0.06  &-  \\ \hline
\end{tabular}

\caption{Material parameters for all the samples. Since the SO fields in the GaAs samples were very small, the SO parameters $\alpha$, $\beta$, and $r$ could not be determined. Furthermore, as the absorption of the GaAs epilayers cannot be measured, the conversion between Faraday angle and spin density cannot be calculated. }
\end{table*}

Spin drag measurements are performed with the electric field applied parallel to the external magnetic field along either the [110] or [1$\overline{1}$0] crystal axes and the time delay fixed to $\Delta t = 13$ ns. The drift velocity $v_d$ is determined from the pump-probe spatial separation at the position with maximum $A_0(x)$ (Fig. \ref{fig1}a,b).  Along these two crystal axes, with this configuration of parallel electric and magnetic fields, the SO field is purely perpendicular to the external magnetic field and manifests as a reduction of the amplitude of the center peak of the magnetic field scans (Fig. \ref{fig1}c). We measure the magnitude of the SO field as a function of applied voltage. 

\begin{figure}
\includegraphics[width=8.5cm]{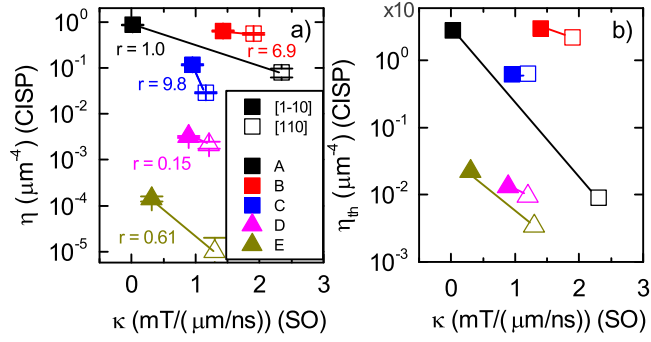}
\caption{\label{fig3} (a) $\eta$ (CISP) vs. $\kappa$ (SO splitting) for all five InGaAs samples. Squares indicate samples with higher indium concentration (2.4\%-2.6\%) and triangles indicate samples with 2.0\% indium. Filled in and open symbols are for measurements along the  $[1\overline{1}0]$ and [110] crystal axes respectively. $r = \alpha/\beta$ characterizes the anisotropy of the SO field. There was a negative differential relationship observed between the two parameters in all five samples. (b) Theoretical calculations for $\eta$ based on the model (Eq. \ref{eq:Raimondi}) using the material parameters for the five InGaAs samples. The model predicts the observed negative differential relationship.}
\end{figure}

The SO field is found to be linear with drift velocity (Fig. \ref{fig1}d), where the slope $\kappa$ is used to characterize the strength of the SO field. Measurements of $\kappa$ for voltages along the [110] and [1$\overline{1}$0] crystal axes allow us to extract the SO parameters $\alpha$ and $\beta$ (Table 1). 
 
CISP is measured with the Faraday rotation of the probe beam in the absence of optical pumping (Fig. \ref{fig2}a). This is described by the equation \cite{Kato_2004}
\begin{equation}\label{CISPeq}
\theta_F = \theta_{\text{el}} \frac{\omega_L\tau}{(\omega_L\tau)^2+1}
\end{equation}
where $\theta_{\text{el}}$ is the amplitude of the electrically induced Faraday rotation, $\omega_L$ is the Larmor precession frequency, and $\tau$ is the transverse spin lifetime. The electrical induced spin density can be related to the electrically induced Faraday rotation with the equation (see Supplemental Material)
\begin{equation}
\rho_{\text{el}} = \frac{\theta_{\text{el}}\rho_{\text{op}}}{2\theta_{\text{op}}}
\end{equation}
where $\rho_{\text{op}}$ and $\theta_{\text{op}}$ are the optically induced spin density and Faraday rotation respectively. The ratio $\rho_{\text{el}}/\theta_{\text{el}}$ for the InGaAs samples is shown in Table 1. The quantity of interest is the density of spins oriented per unit time, given by $\gamma = \rho_{\text{el}}/\tau$. 

The measurement shown in Fig. \ref{fig2} is performed for various voltages applied parallel to the external magnetic field. Fit values for $\rho_{\text{el}}$ and $\tau$ are shown in Fig. \ref{fig2}b,c as a function of the voltage, which is given in terms of the drift velocity. $\gamma$ is found to be proportional to the drift velocity (Fig. \ref{fig2}d), and the slope $\eta$ is used to characterize the electrical spin generation efficiency. Measurements are repeated for voltages along the [110] and [1$\overline{1}$0] crystal axes.

Figure \ref{fig3}a shows the parameter $\eta$ for CISP versus the parameter $\kappa$ for the SO fields for the InGaAs samples. A theory of the inverse spin galvanic effect solely based on the inclusion of intrinsic SO contributions would predict that the CISP should be proportional to the SO field. However, consistent with previous measurements \cite{Norman_2014}, we found that the crystal axis with the smallest SO splitting had the largest CISP and vice versa.


Samples with higher carrier concentrations were found to have greater CISP (Fig. \ref{fig4}a). Assuming the same rate of spin polarization, this would result in a larger spin density given a larger carrier concentration. Furthermore, samples with lower mobility had greater CISP (Fig. \ref{fig4}b). Since the mobility is proportional to the momentum scattering time, this indicates that samples with less time between scattering events had greater spin polarizations, and suggests that an extrinsic polarization mechanism dominates. 

We also found that samples with higher indium concentration had higher electrical spin generation efficiencies (see Fig. S1a). Higher indium concentration causes more strain in the InGaAs epilayer due the 7\% lattice mismatch between InAs and the GaAs substrate. The higher strain results in larger SO splitting in the epilayer. Thus, this suggests that the amount of SO splitting is related to the amount of CISP, albeit not in the direct way described by the model with only Rashba and Dresselhaus SO contributions. There was no clear correlation between the spin dephasing time and the magnitude of CISP (see Fig. S1b). 

In contrast to InGaAs grown on GaAs substrates, GaAs epilayers do not have strain induced spin-orbit fields. However, we also observed CISP in GaAs (see Fig. S2). As with the InGaAs samples, we found that the CISP was greater along the [1$\overline{1}$0] axis than the [110] axis. Furthermore, we found that the GaAs sample with higher carrier concentration had more CISP, consistent with the measurements in InGaAs.

The SO fields in the GaAs samples were very small ($<$0.1 mT) for both the [110] and [1$\overline{1}$0] crystal axis. Since we were able to detect CISP despite the absence of measurable SO fields, this again suggests that the electrically generated spin polarization mechanism is not fully explained with the model with purely intrinsic SO contributions.

Gorini \textit{et al.} derived the Bloch equation for a 2DEG including both intrinsic and extrinsic SO contributions to the spin dephasing, the spin Hall effect, and the spin-generation torque \cite{Gorini_2017}. The change in the total spin polarization over time is given as:
\begin{equation}\label{eq:Raimondi}
\begin{split}
\frac{\partial \vec{S}}{\partial t} = &-(\Gamma_{\text{DP}}+\Gamma_{\text{EY}})\left(\vec{S}-\frac{N_0}{2}\vec{B}_{\text{ext}}\right) - (\vec{B}_{\text{ext}}+\vec{B}_{\text{SO}})\times\vec{S}\\
 &+(\Gamma_{\text{DP}}-\Gamma_{\text{EY}})\frac{N_0}{2}\vec{B}_{\text{SO}}+\frac{\theta_{\text{SH}}^{\text{ext}}}{\theta_{\text{SH}}^{\text{int}}}\Gamma_{\text{DP}}\frac{N_0}{2}\vec{B}_\text{SO}
\end{split}
\end{equation}
where $N_0$ is the density of states, and $\theta_{\text{SH}}^{\text{int(ext)}}$ is the spin Hall angle due to intrinsic (extrinsic) contributions \cite{Engel_2005, Gorini_2010}. $\Gamma_{\text{DP}}$ and $\Gamma_{\text{EY}}$ are the dephasing rate tensors for the two dominant mechanisms: D'yakonov-Perel' (DP) dephasing \cite{Dyakanov_1971}, an intrinsic effect that is due to precession of the spins about momentum-dependent spin-orbit fields between scattering events, and Elliot-Yafet (EY) dephasing \cite{Elliot_1954}, an extrinsic effect that is due to spin flips at scattering events \cite{Raimondi_2009}. 

The relative strength of the DP and EY dephasing mechanisms can be determined from temperature dependent measurements of the spin dephasing time and mobility (see Supplemental Material). At 30 K, the temperature at which all CISP and SO field measurements were performed, the extrinsic EY dephasing mechanism was found to be comparable to or dominant over the intrinsic DP dephasing mechanism for all samples. 

\begin{figure}
\includegraphics[width=8.5cm]{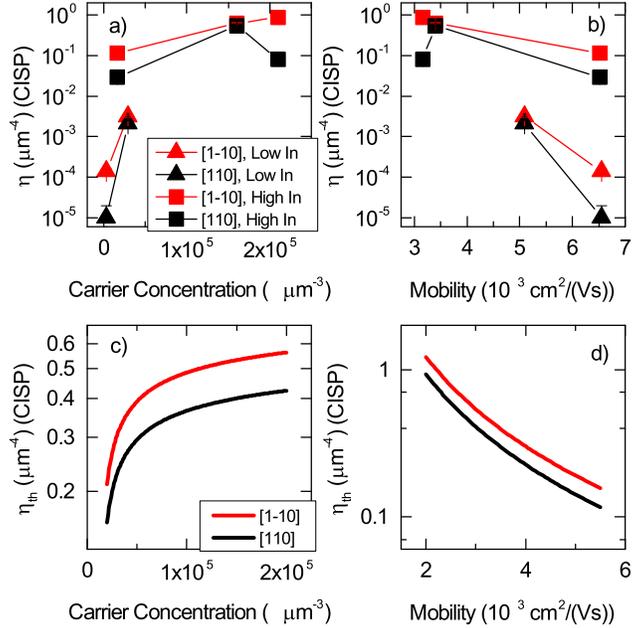}
\caption{\label{fig4} Measured values of $\eta$ (CISP) for the [1$\overline{1}$0] and [110] crystal axes as a function of (a) carrier concentration and (b) mobility. Squares indicate samples with higher indium concentration (2.4\%-2.6\%) and triangles indicate samples with 2.0\% indium. Red and black symbols are for measurements along the  $[1\overline{1}0]$ and [110] crystal axes respectively. Calculations for $\eta$ as a function of (c) carrier concentration and (d) mobility using material parameters for Sample D.}
\end{figure}

Using Eq. \ref{eq:Raimondi}, we can solve for the theoretical steady-state spin density $\rho_{\text{el,th}}$, and therefore the theoretical spin generation rate per unit drift velocity $\eta_{\text{th}}$. The values for $\eta_{\text{th}}$ calculated using the material parameters of the five InGaAs samples are shown in Fig. \ref{fig3}b as a function of the SO splitting along the [1$\overline{1}$0] and [110] crystal axes. For the given material parameters, the model predicts a negative differential relationship between the CISP and SO splitting. In general, the relationship between the CISP and SO splitting may be either positive or negative depending on the values of the spin Hall angles, $r$, and $q$ (see Supplemental Material). Although the predicted values are an order of magnitude larger than the measured values, the relative magnitudes of the predicted $\eta_{\text{th}}$ are qualitatively consistent with the experimental results.

Figure \ref{fig4}c,d, shows $\eta_{\text{th}}$ as a function of carrier concentration and mobility respectively, using the material parameters for Sample D.  The model predicts that the CISP is largest in samples with high carrier concentrations and low mobilities, consistent with the experimental results. 

We performed measurements of CISP and SO splitting along the [110] and [1$\overline{1}$0] crystal axes in seven In$_x$Ga$_{1-x}$As samples with different Indium concentrations and doping densities. In all samples, we found a negative differential relationship between the magnitude of the CISP and SO splitting. Theoretical calculations based on the model proposed by Gorini \textit{et al.} are found to qualitatively agree with the experimental results. This model was derived for a 2DEG, whereas measurements were performed on bulk epilayers. A model that includes 3-dimensional effects may provide better quantitative agreement between the model and the experiment. 

Work by M.L.-K. and V.S. was supported by the U.S. Department of Energy, Office of Basic Energy Sciences, Division of Materials Sciences and Engineering under Award DE-SC0016206. R.R. acknowledges stimulating discussions with C. Gorini, A. Maleki, K. Shen, I. Tokatly, and G. Vignale. S.H., J.O., and R.S.G. were supported in part by the National Science Foundation (Grant No. DMR 1410282); D.D.G was supported in part by the National Science Foundation (Grant No. ECCS 1610362).  

\newcommand{\hbAppendixPrefix}{S}
\renewcommand{\thefigure}{\hbAppendixPrefix\arabic{figure}}
\setcounter{figure}{0}
\renewcommand{\thetable}{\hbAppendixPrefix\arabic{table}} 
\setcounter{table}{0}
\renewcommand{\theequation}{\hbAppendixPrefix\arabic{equation}} 
\setcounter{equation}{0}

\section{Supplementary Information}


\subsection{Converting Faraday Angle to a Spin Density}

It is possible to convert the electrically induced Faraday rotation $\theta_{\text{el}}$ to a spin density by comparing the Faraday rotation due to optical polarization to the Faraday rotation due to electrical polarization \cite{Kato_2004}. With optical injection, the number of spins polarized per laser pulse is 
\begin{equation}
n_{\text{op}} = \rho_{\text{op}} \times 2\pi\sigma_x\sigma_y d = \frac{1}{2}\alpha\left(\frac{P_{\text{pump}}}{f_{\text{rep}}}/2\pi\hbar \frac{c}{\lambda}\right) 
\end{equation}
where $\rho_{\text{op}}$ is the density of optically polarized spins, $\sigma_x$ and $\sigma_y$ are the widths of the Gaussian profile of the pump spot, $d$ is the thickness of the epilayer, $P_{\text{pump}}$ is power of the pump spot, $\alpha$ is the absorption of the epilayer, and $f_{\text{rep}}$ and $\lambda$ are the repetition rate and wavelength of the laser. 

The Faraday rotation due to optical injection is given by
\begin{equation}\label{thetaop}
\theta_{\text{op}} = Ad\iint\left[\rho_{\text{op}}e^{-2\left(\frac{x^2}{2\sigma_x^2}+\frac{y^2}{2\sigma_y^2}\right)}\right] dx dy = \pi A d \rho_{\text{op}} \sigma_x \sigma_y
\end{equation}
where the factor of two in the exponent accounts for the spatial profiles of the pump and probe beams. In this way, we can get the conversion factor $A = \theta_{\text{op}}/(\pi d \rho_{\text{op}}\sigma_x\sigma_y)$ between the spin density and the Faraday rotation. 

The density of electrically polarized spins $\rho_{\text{el}}$ is related to the Faraday rotation by the equation
\begin{equation}\label{thetael}
\theta_{\text{el}} = A d \iint \left[\rho_{\text{el}} \times e^{-\left(\frac{x^2}{2\sigma_x^2}+\frac{y^2}{2\sigma_y^2}\right)}\right] dx dy = 2\pi A d \rho_{\text{el}}\sigma_x\sigma_y
\end{equation}
The electrically induced spin polarization is spatially uniform, and so the only spatial dependence comes from the spatial profile of the probe beam. By including the results for the proportionality factor $A$ into Eq. \ref{thetael}, we arrive at the result for the steady-state density $\rho_{\text{el}}$:
\begin{equation}
\rho_{\text{el}} = \frac{\theta_{\text{el}}\rho_{\text{op}}}{2\theta_{\text{op}}}
\end{equation}
The ratio of the electrically generated spin density $\rho_{\text{el}}$ to the Faraday angle $\theta_{\text{el}}$ for the measurements along both crystal axes of each sample are given in Table 1. 


\subsection{Current Induced Spin Polarization vs. In concentration and T$_2^*$}

\begin{figure}
\includegraphics[width=8.5cm]{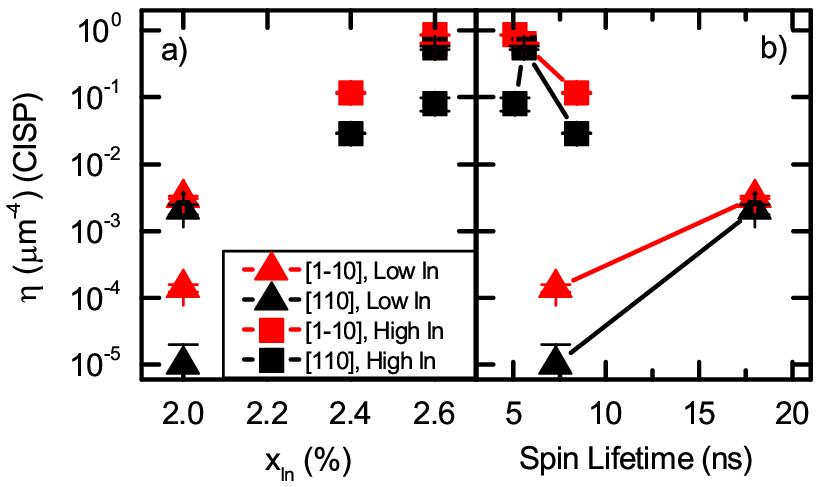}
\caption{\label{figS3} $\eta$ (CISP) for the [1$\overline{1}$0] (red) and [110] (black) crystal axes as a function of (a) indium concentration and (b) spin dephasing time. Square and triangles indicated samples with higher (2.4-2.6\%) and lower (2.0\%) indium concentrations respectively.}
\end{figure}

Samples with higher indium concentration had higher electrical spin generation efficiencies (Fig. \ref{figS3}a). There does not seem to be a relationship between the spin generation efficiency and the spin dephasing time (Fig. \ref{figS3}b).

\subsection{Current Induced Spin Polarization in GaAs epilayers}

Fig. \ref{fig4}a shows CISP for 2 V applied across the higher doped GaAs sample along both the [1$\overline{1}$0] and the [110] crystal axes. Fig. \ref{fig4}b shows CISP for 4 V applied across the lower doped GaAs sample along the [1$\overline{1}$0] and [110] crystal axes. There was no measurable CISP along the [110] axis in the lower doped GaAs sample. In the GaAs samples, the epilayer and substrate absorb at similar wavelengths, and so we cannot measure the absorption of only the epilayer in order to calculate the conversion factor $A$. Because of this, the CISP in the GaAs samples can only be reported in terms of $\mu$rad and the magnitude of CISP in these samples cannot be directly compared to the InGaAs samples.  

\begin{figure}
\includegraphics[width=8.5cm]{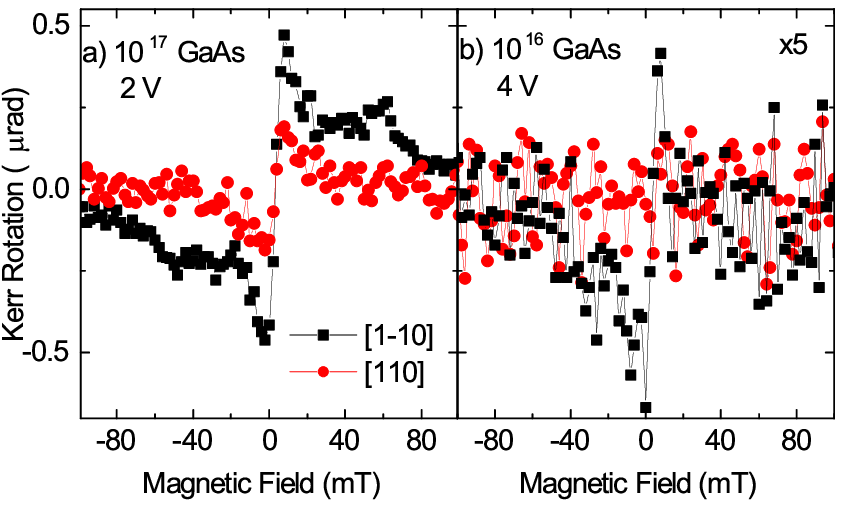}
\caption{\label{fig4} CISP for (a) 2 V across the $n\sim10^{17}$ and (b) 4 V across the $n\sim10^{16}$ (b) GaAs samples. The data in (b) has been magnified by a factor of 5. Black squares and red circles indicate measurements along the $[1\overline{1}0]$ and [110] crystal axes respectively. There was no CISP detected in the  $n=10^{16}$ GaAs sample along the [110] crystal axis. For both samples, CISP was greater along the $[1\overline{1}0]$ crystal axis. Furthermore, the sample with higher carrier concentration had greater CISP.}
\end{figure}


\subsection{Determining the Relative Strength of the Spin Dephasing Mechanisms}

The total spin dephasing time is given by \cite{KikkawaThesis}
\begin{equation}\label{tau}
\frac{1}{\tau_s} = \frac{1}{\tau_{\text{EY}}} + \frac{1}{\tau_{\text{DP}}} = C_{\text{EY}}\mu^{-1}T^2 + C_{\text{DP}}\mu T^3
\end{equation} 
where $T$ is the temperature, and $C_{\text{EY}}$ and $C_{\text{DP}}$ are coefficients denoting the relative strength of the EY and DP dephasing mechanisms.

We performed temperature dependent measurements of the spin dephasing time $\tau_s$ and mobility $\mu$, in order to extract the relative strength of the EY and DP dephasing mechanisms.  

We can fit for $C_{\text{EY}}$ and $C_{\text{DP}}$ in Eq. \ref{tau} using a two independent variable fit, with $T$ and $\mu$ as the independent variables. The relative strength of the two dephasing mechanisms is then defined by 
\begin{equation}\label{qeq}
q(T) = \frac{\tau^{-1}_{\text{EY}}}{\tau^{-1}_{\text{DP}}} = \frac{C_{\text{EY}}}{C_{\text{DP}}}\mu^{-2}T^{-1}
\end{equation}

\begin{figure}
\includegraphics[width=8.5cm]{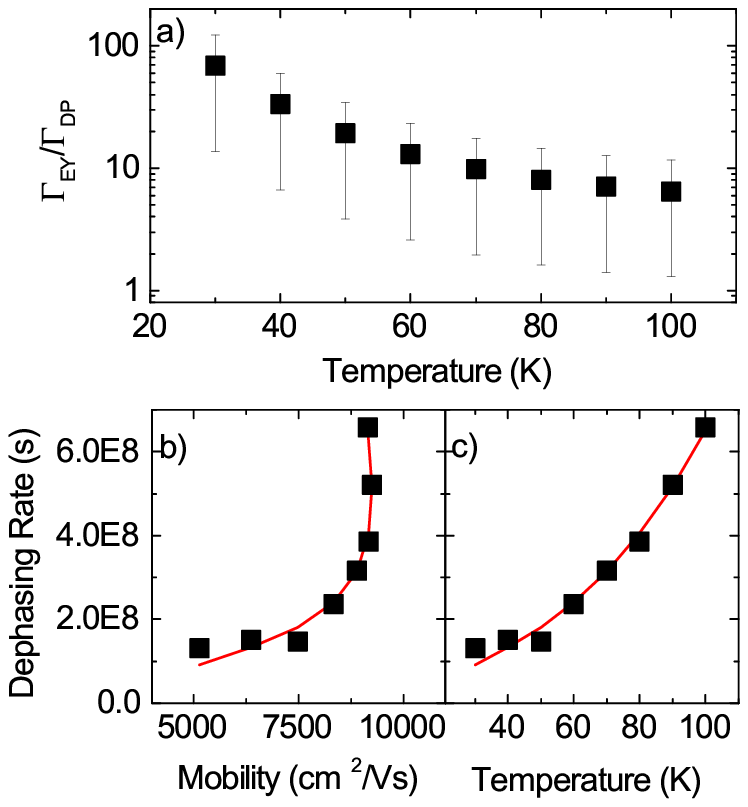}
\caption{\label{fig5} (a) Ratio of the dephasing rates due to the EY and DP dephasing mechanisms for the In$_{0.024}$Ga$_{0.976}$As sample with $n = 1.58 \times 10^{16}$ cm$^{-3}$, calculated from results of a two-independent variable fit of the dephasing rate as a function of the temperature and the temperature-dependent mobility. $q > 1$ indicates that the extrinsic EY dephasing mechanisms dominates at 30 K.}
\end{figure}

The value of $q(T)$ calculated from the fits of the dephasing time and Eq. \ref{qeq} is shown in Fig. \ref{fig5}a. The fits of the dephasing time versus mobility and temperature are shown in Fig. \ref{fig5}b,c for the sample with n = $1.6 \times 10^{16}$ cm$^{-3}$. 


\subsection{Theoretical Spin Generation Efficiency vs $r$ and $q$}

\begin{figure}
\includegraphics[width=8.5cm]{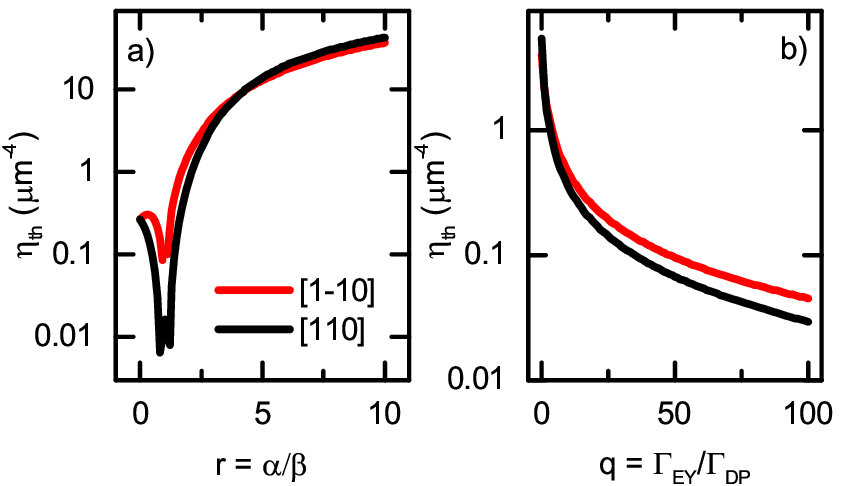}
\caption{\label{figS4} Calculations for $\eta$ as a function of (a) $r = \alpha/\beta$ and (b) $q = \Gamma_{\text{EY}}/\Gamma_{\text{DP}}$ using material parameters for Sample D. The red (black) curve denotes calculations for the [1$\overline{1}$0] ([110]) crystal axis.}
\end{figure}

The theoretical value for $\eta_{\text{th}}$ was also calculated as a function of $r$ and $q$ using material parameters for Sample D. Although the anisotropy in the spin generation efficiency is greatest at $r = 1$, the total magnitude of the efficiency is lowest for that value (Fig. \ref{figS4}a). Although it is necessary to include the EY dephasing mechanism in order to get the negative differential relationship between CISP and the SO splitting, samples in which the DP dephasing mechanism dominates (i.e. $q < 1$) have larger electrical spin generation efficiencies (Fig. \ref{figS4}b).  

\subsection{Crystal Axis with Maximum CISP}

For the materials under study in this work, CISP was measured to be maximum along the [1$\overline{1}$0] crystal axis. However, depending on the values of the spin Hall angles, $r$, and $q$, the model predicts maximum spin polarizations along either the [1$\overline{1}$0] or the [110] crystal axis. 

We begin by defining the following dimensionless parameters (in units of $\hbar = c = 1$):
\begin{equation}
\begin{split}
b &= 2e\tau\beta E \tau_{\text{DP}} \\
s_{\text{H}} &= \frac{\theta_{\text{SH}}^{\text{ext}}+\theta_{\text{SH}}^{\text{int}}}{\theta_{\text{SH}}^{\text{int}}},
\end{split}
\end{equation}
the dimensionless matrices
\begin{equation}
\begin{split}
\hat{\gamma}_{\text{tot}} &=\left( \begin{matrix}
1+r^2+q & 2r & 0 \\
2r & 1+r^2 +q & 0 \\
0 & 0 & 2(1+r^2)
\end{matrix} \right) \\
\hat{\gamma}_{\text{rel}} &=\left( \begin{matrix}
s_{\text{H}}(1+r^2)-q & 2s_{\text{H}}r & 0\\
 2s_{\text{H}}r & s_{\text{H}}(1+r^2)-q & 0 \\
0 & 0 & 2s_{\text{H}}(1+r^2)
\end{matrix} \right),
\end{split}
\end{equation}
and the dimensionless vector
\begin{equation}
\vec{b} = \left(\begin{matrix}
\hat{E}_x + r\hat{E}_y \\
-(r\hat{E}_x + \hat{E}_y)\\
0
\end{matrix}\right),
\end{equation}
in the \{[100], [010], [001]\} basis.

In the absence of an external magnetic field, the Bloch equation for the model proposed by Gorini \text{et al.} \cite{Gorini_2017} can be written as
\begin{equation}
\partial_t\vec{S}=-\hat{\gamma}_{\text{tot}}(\vec{S}-b\hat{\gamma}_{\text{tot}}^{-1}\hat{\gamma}_{\text{rel}}\vec{b}) - b\vec{b} \times \vec{S}
\end{equation}

The steady-state solution for the in-plane spin polarization has the compact form
\begin{equation}
\left(\hat{\mu}+\frac{b^2}{2(1+r^2)}\hat{\omega}\right)\vec{S}_{xy} = b\hat{\nu}\vec{b}_{xy}
\end{equation}
where
\begin{equation}
\begin{split}
\hat{\mu}&=\left(\begin{matrix}
1+r^2+q & 2r \\
2r & 1+r^2+q
\end{matrix}\right) \\
\hat{\omega}&=\left(\begin{matrix}
b_y^2 & -b_x b_y \\
-b_x b_y & b_x^2
\end{matrix}\right) \\
\hat{\nu}&=\left(\begin{matrix}
s_{\text{H}}(1+r^2)-q & 2s_{\text{H}}r \\
2s_{\text{H}}r  & s_{\text{H}}(1+r^2)-q
\end{matrix}\right)
\end{split}
\end{equation}

To first order in $b$, the steady-state in-plane spin polarization in thus given by
\begin{equation}
\vec{S}_{xy}^{(1)}=b\hat{\mu}^{-1}\hat{\nu}\vec{b} 
\end{equation}

This can also be written in the form
\begin{equation}
\vec{S}_{xy}^{(1)} = (a_0\hat{\sigma}^0+a_x\hat{\sigma}^x)\vec{b}
\end{equation}
where $\hat{\sigma}^0$ is the identity, $\hat{\sigma}^x$ is the Pauli matrix, and
\begin{equation}
\begin{split}
a_0 &= \frac{b}{(1+r^2+q)^2-4r^2}\left[s_{\text{H}}(1-r^2)^2+q(s_{\text{H}}-1)(1+r^2)-q^2\right] \\
a_x &= \frac{2rq(s_{\text{H}}+1)b}{(1+r^2+q)^2-4r^2}
\end{split}
\end{equation}

When $\vec{E}\parallel[110]$, $\vec{b}\parallel[1\overline{1}0]$ and $\vec{S}_{xy} = (a_0-a_x)\vec{b}$, and when $\vec{E}\parallel[1\overline{1}0]$, $\vec{b}\parallel[110]$ and $\vec{S}_{xy} = (a_0+a_x)\vec{b}$. 

If $s_{\text{H}} + 1>0$, then $a_x>0$. This happens as long as $\theta_{\text{SH}}^{\text{ext}} / \theta_{\text{SH}}^{\text{int}}>-2$. Therefore, the crystal axis with the maximum steady-state spin polarization is dependent on the sign of $a_0$. $a_0$ vanishes when $s_{\text{H}}=\frac{q(1+r^2+q)}{(1-r^2)^2+q(1+r^2)}$, and so we have
\begin{equation}
\begin{split}
s_{\text{H}}&>\frac{q(1+r^2+q)}{(1-r^2)^2+q(1+r^2)} \text{: $a_0>0$ and $\vec{S}_{xy}$ maximum for $\vec{E}\parallel[1\overline{1}0]$} \\
s_{\text{H}}&<\frac{q(1+r^2+q)}{(1-r^2)^2+q(1+r^2)} \text{: $a_0<0$ and $\vec{S}_{xy}$ maximum for $\vec{E}\parallel[110]$} 
\end{split}
\end{equation}

\end{document}